\documentclass[aps,prd,10pt,onecolumn,superscriptaddress,nofootinbib,showkeys,showpacs,altaffilletter]{revtex4-1}

\usepackage{graphicx}
\usepackage{dcolumn}
\usepackage{multirow}
\usepackage{amssymb}
\usepackage{amsmath}
\usepackage{amsfonts}
\usepackage{amsbsy}
\usepackage{color}
\usepackage{rotating}
\usepackage[english]{babel}

\newcommand{\be}{\begin{equation}}
\newcommand{\ee}{\end{equation}}
\newcommand{\bea}{\begin{eqnarray}}
\newcommand{\eea}{\end{eqnarray}}

\def \omms   {\Omega_m}

\renewcommand{\(}{\left(}

\newcommand{\lcdm }{$\Lambda$CDM }

\begin{document}

%\title{Should, in view of the forthcomming surveys, scale-invariant aproach to the perturbations, be abandoned finally?}
%\title{Are we in the epoch of precision cosmology, when the scale invariance of the perturbations should be abandoned finally?}

% maybe (?) my favourite
\title{Scale-dependent perturbations finally detectable by future galaxy surveys and their contribution to cosmological model selection}

\date{\today}

\author{Tomasz Denkiewicz}
\email{tomasz.denkiewicz@usz.edu.pl}
\affiliation{\it Institute of Physics, Faculty of Mathematics and Physics, University of Szczecin, Wielkopolska 15,
          70-451 Szczecin, Poland}
\affiliation{\it Copernicus Center for Interdisciplinary Studies, S{\l }awkowska 17, 31-016 Krak\'ow, Poland}
\author{and Vincenzo Salzano}
\email{vincenzo.salzano@usz.edu.pl}
\affiliation{\it Institute of Physics, Faculty of Mathematics and Physics, University of Szczecin, Wielkopolska 15,
          70-451 Szczecin, Poland}

\begin{abstract}
The question of the origin of the present accelerated expansion of the Universe is still pending. By means of the present geometrical and dynamical observational data, it is very hard to establish, from a statistical perspective, a clear preference among the vast majority of the proposed models for the dynamical dark energy and/or modified gravity theories alternative with respect to the $\Lambda$CDM scenario. On the other hand, on scales much smaller than present Hubble scale, there are possibly detectable differences in the growth of the matter perturbations for different modes of the perturbations, even in the context of the $\Lambda$CDM model. In view of the new planned observations, that will give insight into the perturbations of the dark sector, this issue is being worth of further investigation. Here, we analyze the evolution of the dark matter perturbations in the context of $\Lambda$CDM and some dynamical dark energy models involving future cosmological singularities, such as the sudden future singularity and the finite scale factor singularity. We employ the scale-dependent perturbation equations for the growth function, and we abandon both the sub-Hubble approximation and the slowly varying potential assumption, which lead to the well known and most commonly used scale independent solutions for the perturbations. We apply the Fisher Matrix approach to three future planned galaxy surveys e.g., DESI, \textit{Euclid}, and \textit{WFirst-2.4}, in order to have insights on the possibility to confute cosmological models through perturbations growth data in the next future. With the mentioned surveys on hand, only with the dynamical probes, we will achieve multiple goals: $1.$ the improvement in the accuracy of the determination of the $f\sigma_{8}$ will give the possibility to discriminate between the $\Lambda$CDM and the alternative dark energy models even in the scale-independent approach; $2.$ it will be possible to test the goodness of the scale-independence finally, and also to quantify the necessity of a scale dependent approach to the growth of the perturbations, in particular using surveys, which encompass redshift bins with scales $k<0.005\,h$ Mpc$^{-1}$; $3.$ the scale-dependence itself might add much more discriminating power in general, but further advanced surveys will be needed.
\end{abstract}

\maketitle

\section{Introduction}
\setcounter{equation}{0}

Current observations on cosmological scales, like the measurements of the luminosity distance with the Type Ia Supernovae (SNeIa) \cite{1998AJ,1999ApJ,Suzuki:2011hu,Betoule:2012an}, the clustering scale of galaxies by detection of the baryon acoustic oscillations (BAO) \cite{Eisenstein,Bassett:2009mm,Alam2016}, and the scaled distances to the last scattering surface ($\mathcal{R}, l_a$) \cite{9702100v2,Wang:2007mza,Wang:2015tua}, are all geometrical probes. While, on one hand, they have helped to state the existence of dark energy, are useful probes to quantify the dark energy amount, and give information about the background expansion history ($H(z)$) \cite{Weinberg:2012es,Salzano:2013jia,Lazkoz:2013ija}, on the other hand, they have also revealed to be quite unable to gain some more deep insight into the nature of dark energy. In fact, with the geometrical probes, we are unable to distinguish, in the context of general relativity, among models for which the accelerated expansion is driven by a cosmological constant ($\Lambda$CDM scenario) or a dynamical fluid; or, in a more general theoretical context, to discriminate between general relativity and other possible modified gravity theories.

Other types of complementary tests are the dynamical probes, which allow to probe the growth of the matter perturbations. The data up to now do not provide very restrictive constraints onto many cosmological parameters, and for many of the models which are currently on the market (see, for example, \cite{Ade:2015xua,Alam:2015rsa,Albarran:2016mdu} and references therein); but the combination of geometrical probes and the density contrast $\delta_m(z)=\delta \rho_m(z)/\rho_m(z)$ give much more power to differentiate between the dark energy and the modified gravity models (such as $f(R)$ models and Gvali-Gabadadze-Porrati (DGP) model), and the $\Lambda$CDM \cite{0908.2669v1,0809.3374v2,astro-ph/0701317v2,Bamba:2012qi} than geometrical or dynamical probes alone. The lack of a firm explanation of the origin of the present universe accelerated expansion led to the formulation of different scenarios of modified gravity models or dynamical dark energy models among which, within general relativity, some foresaw different types of singularities in a finite future time. In this latter field, models describing a Big-Rip (BR), a Sudden Future Singularity (SFS), also known as Type II singularities \cite{Barrow:2004xh, Barrow:2004gk,Dabrowski:2004bz,hep-th/0505215v4}, a Generalized Sudden Future Singularity  (GSFS), a Finite Scale Factor Singularity (FSF), also known as Type III singularities \cite{0910.0023v1,hep-th/0505215v4}, a Big-Separation Singularity (BS) \cite{Nojiri:2005sr,hep-th/0505215v4} and $w$-singularities \cite{0902.3107v3} gained some attention among cosmologists.

Here, we will deal with the possibility of testing observationally, constraining, differentiating and hopefully falsifying some of those scenarios using forthcoming data from planned surveys which intend to give constraints onto the growth function \cite{astro-ph/0703191v2}. Future galaxy surveys like DESI\footnote{\url{http://desi.lbl.gov/}} \cite{Levi:2013gra}, \textit{Euclid}\footnote{\url{http://sci.esa.int/euclid/}} \cite{Cimatti:2009is,Refregier:2010ss,Laureijs:2011gra,Amendola:2016saw} and \textit{WFirst-2.4}\footnote{\url{http://wfirst.gsfc.nasa.gov/.}} \cite{Spergel:2013tha} should be potentially able to discriminate between the cosmological constant and evolving dark energy scenarios. The forthcoming observations of the cosmic background radiation in the microwave to far-infrared bands in the polarization and the amplitude, as the Polarized Radiation and Imaging Spectroscopy (PRISM) \cite{prism} and the very high precision measurements of the polarization of the microwave sky by the Cosmic Origins Explorer (CoRE) satellite \cite{core}, should also improve the constraints onto the dark sector \cite{Errard:2015cxa,DiValentino:2016foa}, and help to further settle this topic.

In this work we go beyond the scale-independent approximation for the equation for the matter density evolution, which is nowadays most commonly applied, and consider the full set of the scale dependent equations. We show how the sensitivity reached by the considered future surveys will be able to give a first and maybe definitive answer about the validity of the scale-independent approach. Moreover, we show that it will be possible to discriminate between different models of dynamical dark energy, which foresee different kinds of singularities in the course of the evolution of the universe to appear. The evolution of the linear density perturbations for FSFS, Big Rip, SFS, FSFS and Pseudo-Rip was a subject of some previous works \cite{1202.3280v2,AO,1411.6169v2} (similar analysis for other singularity models are also in \cite{Albarran:2016mdu}).

In the following Sec.~\ref{perturbations} we introduce the background equations, and the perturbed Einstein equations in the Newtonian frame, which are later translated into the Synchronous gauge \cite{astro-ph/9506072v1} for the purpose of making the comparison of the derived power spectra with the surveys of concern. In Sec.~\ref{themodels} we give a short description of the dynamical dark energy models with the SFS and the FSFS. In Sec.~\ref{forecast} we describe in detail the Fisher Matrix approach we have applied to the selected future galaxy surveys. Finally, in Sec.~\ref{lastS}, we describe the results for the dark matter perturbation evolution and the growth function, in $\Lambda$CDM and in models with singularities, for different scales of the perturbation modes and for different surveys.

\section{Scale dependent growth function}\label{perturbations}
\setcounter{equation}{0}
The growth function is a useful tool, as a probe of the dynamics of the universe expansion; it is often quantified through the growth rate $f$ defined as
\be
f(a)\equiv \frac{d\ln \delta_{m}(a)}{d\ln a}, \label{fofa}
\ee
which is, on turn, parameterized by the expression:
\be
f(a)=\omms(a)^\gamma, \label{fofa2}
\ee
where $a$ is the scale factor, $\gamma$ is the growth index, and $\Omega_m(a)$ the dimensionless matter density parameter function:
\be
\Omega_{m}(a) \equiv \frac{H_0^2 \Omega_{m,0} a^{-3}}{H(a)^2} \label{omadef},
\ee
with $\Omega_{m,0}$ the dimensionless matter density parameter today, $H(a) = \dot{a}/a$ the expansion rate and $H_0$ the Hubble constant (expansion rate today). This parametrization is an approximation to the scale independent solution of the equation for the growth rate \cite{0808.2689v4,0903.5296v2}:
\be
f' + f^2 + f\left(\frac{\dot H}{H^2}+2\right)=\frac{3}{2} \omms \label{greqlna},
\ee
where $'\equiv \frac{d}{d\ln a}$, which is obtained from the density contrast evolution equation:
\be
{\ddot \delta}_m + 2 H {\dot \delta}_m - 4\pi G \rho_m \delta_m =0, \label{dcesi}
\ee
with the change of the variable from the time $t$ to the scale factor $\ln a$, where an overdot denotes the derivative with respect to the time and $\rho_m$ is the matter density. It was found that $\gamma=6/11$ for the $\Lambda$CDM \cite{astro-ph/9804015v1,astro-ph/0507263v2,astro-ph/0701317v2}, and its best observational determination is $0.665\pm0.0669$ in \cite{Johnson:2015aaa}. For dark energy models with a slowly varying equation of state, the solution is well approximated by Eq.~(\ref{fofa2}) with
\be
\gamma=\frac{3(w_0-1)}{6w_0-5},
\ee
where $w_0\equiv w(a)=p/\rho$ is the dark energy equation of state (for $\Lambda$CDM, $w_0=-1$). It has been shown, that for slowly varying dark energy models, the $\gamma$ index is not strongly time dependent and if it varies, it does only at a few percent level \cite{1002.3030v1,0903.5296v2}. For modified gravity models, this is not longer the case. The $\gamma$ may vary and for the well known case of the DGP model, for example, $\gamma\simeq 0.68$ and a better approximation to the solution of the full set of equations Eqs.~(\ref{gr1})~-~(\ref{gr3}) is obtained, when the redshift dependence is taken into account. Various parameterizations for the growth index have been proposed in \cite{0905.2470v2,astro-ph/0701317v2,0802.4196v4,0808.1316v2,Bamba:2012qi}.

Having in mind the forthcoming data for the growth of the perturbations and their improved accuracy, it seems natural to start question if one should not be careful while using the scale independent approximation during the derivation of Eq.~(\ref{dcesi}) \cite{Alam:2015rsa}. In fact, it has been found \cite{0808.2689v4,0903.5296v2,1002.3030v1}, that the exact solution for the full set of the scale-dependent equations shows a scale dependence on scales larger than $100\,h^{-1}$ Mpc for the $\Lambda$CDM. It has been argued, that the scale invariant approximation breaks down because the sub-Hubble scale assumption for the perturbation modes is used and this one might break down already for scales around $200\,h^{-1}$ Mpc in the early stages, no matter that recently the Hubble scale is around $3000\, h^{-1}$ Mpc. Actually, this is the case for the $\Lambda$CDM and the singularity scenarios which are considered in this work \cite{0808.2689v4,0903.5296v2,1002.3030v1,1411.6169v2}. It was also shown in \cite{1411.6169v2}, that the models of dynamical dark energy with singularities differ with respect to the mode wave-number, for which the amplitude of the dark energy perturbation is of the order of the dark matter perturbation amplitude. It is important to have in mind, that in the most general case in each model there exist limiting wavelength, for which the perturbations in the dark energy couple with the dark matter perturbations and can not be ignored. The two models we have considered here are, of course, two very limited cases of a more general picture. In fact, while for quintessence models the perturbations in the dark energy can play a significant role only on scales comparable to Hubble scale, for models for which the scalar field is not canonical, or for dynamical dark energy models in which dark energy speed of sound is $c_s<<1$, the perturbations in dark energy can grow as perturbations in the dark matter for even smaller scales \cite{0806.3461v1,0909.0007v2,1303.0414v2,1411.6169v2}.

In this work we consider dark matter perturbations solely. In order to test the scale dependent approximation in the singular models in the view of the new forthcoming data, we defy the assumption of the sub-Hubble scale of the perturbation modes. We consider the scale dependent solution to the full set of the perturbation equations, for both the FSFS and SFS models, and we compare them with the $\Lambda$CDM as a reference model with the same general assumptions.

The perturbed metric in the Newtonian gauge, with the assumption of the lack of the anisotropic stress, takes the form:
\begin{equation}
 ds^2 = -(1+2\Phi) dt^2 + (1-2\Phi)a^2\gamma_{ij}dx^i dx^j,
\end{equation}
where $\gamma_{ij}$ is the spatial part of the metric and  $\Phi$ is the Newtonian potential. With the assumption that the universe is flat and filled only with pressureless, non-relativistic dark matter, $\rho_m$, and an exotic fluid (which we call dark energy) with energy density $\rho_{de}$, the evolution is governed by the background Friedmann equations:
\bea H^2 &=& \frac{8\pi G}{3}(\rho_m +\rho_{de}), \label{fried} \\ {\dot \rho} &=& -3H (\rho + p) \label{cont} \eea
and by the perturbed up to linear order Einstein equations in Newtonian gauge, which finally altogether result in the set of equations:
\bea
\label{gr1}\ddot{\Phi}&=&-4H\dot{\Phi}+8\pi G \rho_{de} w_{de}\Phi,\\
\label{gr2}\dot{\delta}&=&3\dot{\Phi}+\frac{k^2}{a^2}v_{f},\\
\label{gr3}\dot{v}_{f}&=&-\Phi,
\eea
with the following constraint equations:
\bea
\label{constraint1}3H(H\Phi+\dot{\Phi})+\frac{k^2}{a^2}\Phi&=&-4\pi G\delta\rho_m,\\
\label{constraint2}(H\Phi+\dot{\Phi})&=&-4\pi G\rho_m v_{f}.
\eea
Here $v_f=-v\,a$, and $v$ is the velocity potential for the dark matter; $k$ is the wavenumber. With the sub-Hubble approximation $k^2/a^2>>H^2$ and a slowly varying gravitational potential one obtains the scale independent equation for the matter density contrast evolution, Eq.~(\ref{dcesi}). When the sub-Hubble approximation is relaxed and the approximation of a slowly varying Newtonian potential is hold one gets the scale dependent evolution for the $\delta_m$ in the following form \cite{1002.3030v1,0903.5296v2}:
\bea {\ddot \delta_m} + 2 H {\dot \delta_m} - \frac{4\pi G \rho_m \delta_m}{1+\xi(a,k)}=0,\label{delsd} \eea
where \bea \xi(a,k)=\frac{3 a^2 H(a)^2}{k^2}. \label{xidef} \eea
Eq.~(\ref{delsd}) may be expressed in terms of the growth factor, $f=\frac{d\ln \delta_m}{d\ln a}$ as
\bea f' + f^2 + \(2-\frac{3}{2} \omms(a)\)f=\frac{3}{2}\frac{\omms(a)}{1+\xi(a,k)},\label{fsd}  \eea
where $'\equiv \frac{d}{d\ln a}$ and assuming \lcdm for $H(a)$. The approximate solution to Eq.~(\ref{fsd}) can be parameterized as following:
\be
f(k,a)=\frac{\omms(a)^\gamma}{1+\frac{3H_0^2\Omega_{m,0}}{ak^2}}.\label{approxfk}
\ee
For the sub-Hubble scales $\xi(k,a)\rightarrow 0$, and Eq.~(\ref{fsd}) reduces to Eq.~(\ref{greqlna}), for which the solution is well approximated by Eq.~(\ref{fofa2}) with the $\gamma=\frac{6}{11}$. It has been shown \cite{1002.3030v1,0903.5296v2} that Eq.~(\ref{delsd}) provides a better approximation to the full general relativistic system Eqs.~(\ref{gr1})~-~(\ref{gr3}) up to the horizon scales, while for larger scales one can not ignore the change in the time of the potential, $\Phi$.

Anyway, it is important to stress, that in this work we will not use any of the two above approximations, i.e., we will not assume nor the sub-Hubble nor slowly varying potential approximation.

%%%%%%%%%%%%%%%%%%%%%%%%%%%%%%%%%%
%%%%%%%%%%%%%%%%%%%%%%%%%%%%%%%%%% SECTION

\section{FSFS and SFS as the dynamical dark energy candidates}\label{themodels}
\setcounter{equation}{0}

The SFS and FSFS show up within the framework of the Einstein-Friedmann cosmology governed by the standard field equations, Eqs.~(\ref{fried}),
and the energy-momentum conservation law, Eq.~(\ref{cont}). We get the SFS and FSFS scenarios with the scale factor in the following form:
\be
\label{sf2} a(t) = a_s \left[b + \left(1 - b \right) \left( \frac{t}{t_s} \right)^m - b \left( 1 - \frac{t}{t_s} \right)^n \right]\;.
\ee
An appropriate choice of the constants $(b, t_s, a_s, m,n)$ is necessary \cite{Barrow:2004xh,DHD}. For both cases, the SFS as well as the FSFS model, the evolution starts with the standard big-bang singularity at $t=0$ $(a=0)$, and evolves to an exotic singularity for $t=t_s$, where $a=a_s\equiv a(t_s)$ is a constant. Accelerated expansion in an SFS universe is assured with a negative $b$, while for a FSFS universe $b$ has to be positive. In order to have the SFS, $n$ has to be within the range $1<n<2$; while for an FSFS, $n$ has to obey the condition $0<n<1$. For the SFS at $t=t_s$, $a \to a_s$, $\varrho \to \varrho_s=$ const., $p \to \infty$; while for an FSFS the energy density $\rho$ also diverges and one has: for $t\rightarrow t_s$, $a\rightarrow a_s$, $\rho\rightarrow\infty$, and $p \rightarrow \infty$, where $a_s,\ t_s$, are constants and $a_s\neq 0$. In both scenarios the non-relativistic matter scales as $a^{-3}$, i.e.
\be
\rho_m=\Omega_{m,0}\rho_0\left(\frac{a_0}{a}\right)^3\; ,
\ee
and the evolution of the exotic (dark energy) fluid $\rho_{de}$, can be determined by taking the difference between the total energy density $\rho$, and the energy density of the non-relativistic matter, i.e.
\be
\rho_{de}=\rho-\rho_m~~.
\ee
In those scenarios the dark energy component is also responsible for the exotic singularity at $t\rightarrow t_s$. The dimensionless energy densities are defined in a standard way as
\be
\Omega_m=\frac{\rho_m}{\rho}, \hspace{0.3cm} \Omega_{de}=\frac{\rho_{de}}{\rho}\; .
\ee
For the dimensionless exotic dark energy density we have the following expression
\be
\label{Omde}
\Omega_{de}=1-\Omega_{m,0}\frac{H_0^2}{H^2(t)}\left(\frac{a_0}{a(t)}\right)^3=1-\Omega_{m}.
\ee
The barotropic index of the equation of state for the dark energy is defined as
\be
\label{wde}
w_{de}=p_{de}/ \rho_{de}.
\ee
The singularity scenarios considered in this work were also tested as candidates for dynamical fine structure cosmology \cite{alpha}. In that approach, the dark energy is sourced by a scalar field which couples to the electromagnetic sector of the theory; the knowledge about the effective evolution of the dark energy density and the dark energy equation of state evolution, is sufficient to estimate the resulting fine structure evolution.

\section{Forecast}\label{forecast}
\setcounter{equation}{0}

In order to explore the forecast power for future galaxy surveys, we employ the Fisher Matrix approach to calculate the expected errors on the observational quantity $f\sigma_{8}$.

\subsection{Fiducial cosmological background}\label{fcb}

First, we have to state the values for the main cosmological parameters, which characterize the fiducial cosmology, at which the Fisher Matrix elements are calculated. We have chosen the \textit{Planck} baseline $\Lambda$CDM model $2.71$ from the \textit{Planck Legacy Archive}\footnote{\url{https://wiki.cosmos.esa.int/planckpla2015/images/f/f7/Baseline_params_table_2015_limit68.pdf}} and reported its parameters in Table~\ref{tab:fiducial_LCDM}.

\begin{center}
{\renewcommand{\tabcolsep}{1.mm}
{\renewcommand{\arraystretch}{1.25}
\begin{table}[h!]
\begin{minipage}{\textwidth}
\centering
\resizebox*{\textwidth}{!}{
\begin{tabular}{cccccccccccccc}
\hline
\hline
$\Omega_{c}h^{2}$ & $\Omega_{b}h^{2}$ & $\Omega_{k}$ & $h$ & $w$ & $n_{s}$ & $r$ & $\ln (A_{s} 10^{-9})$ & $\tau$ & $z_{re}$ & $\Omega_{\nu}h^2$ & $N_{eff}$ & $Y_{He}$ & $\sigma_{8,0}$  \\
\hline
$0.11865$ & $0.022307$ & $0$ & $0.67783$ & $-1$ & $0.96722$ & $0$ & $2.14666$ & $0.0677$ & $8.99$ & $0.00065$ & $3.046$ & $0.246692$ & $0.8163$ \\
\hline
\hline
\end{tabular}}
\caption{Fiducial cosmological model. Note that the value of $\sigma_{8}$ is calculated by CAMB once the other parameters are given.}\label{tab:fiducial_LCDM}
\end{minipage}
\end{table}}}
\end{center}

Other useful cosmological quantities, which are needed to both solve Eqs.~(\ref{gr1})~-~(\ref{constraint2}) and calculate the Fisher Matrices require the Hubble function:
\begin{equation}
H^{2}(a)= H^{2}_{0}\left[\Omega_{m,0} a^{-3} + \Omega_{k} a^{-2} + (1-\Omega_{m,0}-\Omega_{k})\right]  \; ,
\end{equation}
where $\Omega_k = k^2/H_0^2$ is the spatial curvature parameter, and $H_{0} = 100~h$ is the Hubble constant. For distances and volumes we follow the notation in \cite{Hogg:1999ad}.

The matter density contrast is obtained by solving the differential equations Eqs.~(\ref{gr1})~-~(\ref{gr2})~-~(\ref{gr3}) with constraints given by Eqs.~(\ref{constraint1})~-~(\ref{constraint2}). We stress again that no sub-Hubble approximation and no slowly varying potential condition is applied to the above equations in this work. The matter density contrast $\delta_m$ is normalized at $a=1$. Note also that Eqs.~(\ref{gr1})~-~(\ref{gr2})~-~(\ref{gr3}) are in Newtonian gauge so, in order to use properly the matter density contrast in the calculation of the power spectrum, once we solve them, we translate the matter density contrast to the Synchronous gauge by using \citep{Challinor:2011bk,Jeong:2011as,Bruni:2011ta}
\begin{equation}
\delta^{S}_{m}(k,a) = \delta_{m}(k,a) - 3 \frac{a H(a)}{k} v(k,a)\;
\end{equation}
where, of course, the suffix $S$ refers to the Synchronous gauge quantity, and $\delta_{m}$ is the solution in the Newtonian gauge. Once we have $\delta^{S}_{m}(k,a)$, we calculate the other useful quantities, i.e. the amplitude of the (linear) power spectrum on the scale of $8h^{-1}$ Mpc:
\begin{equation}
\sigma_{8}(k,a)=\sigma_{8,0} \delta^{S}_{m}(k,a)\; ,
\end{equation}
where the suffix ``$0$'' means evaluated at present time ($z=0$ or $a=1$); and the linear growth rate
\begin{equation}
f(k,a) = \frac{d\, \ln \delta^{S}_{m}(k,a)}{d\, \ln a}\,, \qquad f_{s}(k,a) = f(k,a)\, \sigma_{8}(k,a) \,.
\end{equation}
The theoretical linear matter power spectrum is calculated using CAMB\footnote{\url{http://camb.info/}} \cite{Lewis:1999bs,Howlett:2012mh,Challinor:2011bk,Lewis:2007kz} outputs for the transfer function $\mathcal{T}$\footnote{We assume that such transfer functions are unchanged also for the singularity models which we consider. Given that singularity should occur in a very late future, and that early evolution is insensitive to it, such assumption is generally very reasonable.} expressed, incidentally, in the Synchronous gauge:
\begin{equation}
\mathcal{P}_{L}(k,a) = A_{s} \left( \frac{k}{k_{piv}}\right)^{n_s} \mathcal{T}(k)^2 \left(\frac{\delta^{S}_{m}(k,a)}{\delta^{S}_m(k,1)}\right)^{2}\; ,
\end{equation}
where the pivot scale $k_{piv}$ is $0.05$ Mpc$^{-1}$, and $A_{s}$ is from the fiducial cosmology.

\subsection{Defining the survey}

We have chosen to focus on three different future galaxy surveys: DESI, \textit{Euclid} and \textit{WFirst-2.4}. For each survey we specify: the redshift range and the bin width, where the galaxy correlation is measured; the total area (in square degrees); the galaxy bias $b_{g}$ and the redshift errors for each galaxy type used as mass tracer; the galaxy number density, generally expressed as $dN / dz dA$, i.e., as the number of galaxies observed per redshift bin and per square degree; the systematic shot noise, defined as $\mathcal{P}_{shot}(z) = 1 / n(z)$. Most of these specifications are given in Table~\ref{tab:surveys}, galaxy number densities are taken from \cite{Font-Ribera:2013rwa}. Note that the galaxy bias function is generally derived in a context in which both the sub-Hubble and the slowly varying potential approximations are assumed. Thus, in order to be able to properly use such galaxy bias functions, we consider the limit $k\rightarrow\infty$ of our solutions, which, as stated above, are calculated without any of the former assumptions, and only in this limit are equal to the standard scale-independent results.

\begin{center}
{\renewcommand{\tabcolsep}{1.5mm}
{\renewcommand{\arraystretch}{1.5}
\begin{table}[t!]
\begin{minipage}{\textwidth}
\centering
\resizebox*{\textwidth}{!}{
\begin{tabular}{ccccccccc}
\hline
\hline
 & $(z_{min},z_{max})$ & $\Delta z$ & $A_{surv}$   & $\sigma_{z}$  & $b_{g}$ & $k^{bin}_{max}$ & $k^{bin}_{min}$ & $k^{survey}_{min}$ \\
 &                     &            & (sq. deg.)   &               &         & ($h$ Mpc$^{-1}$) & ($h$ Mpc$^{-1}$) & ($h$ Mpc$^{-1}$) \\
\hline
\multirow{3}{*}{DESI}     & \multirow{3}{*}{$(0.1,1.9)$}   & \multirow{3}{*}{$0.1$} & \multirow{3}{*}{$14000$} & \multirow{3}{*}{$0.001\ (1+z)$} & ELG: $0.76 \frac{\delta_{m}(\infty,z)}{\delta_{m}(\infty,0)}$ & \multirow{3}{*}{$1.0 \cdot 10^{-2}$} & \multirow{3}{*}{$3.6 \cdot 10^{-3}$} & \multirow{3}{*}{$1.6 \cdot 10^{-3}$}\\
& & & & & LRG: $1.7 \frac{\delta_{m}(\infty,z)}{\delta_{m}(\infty,0)}$ & & & \\
& & & & & QSO: $1.2 \frac{\delta_{m}(\infty,z)}{\delta_{m}(\infty,0)}$ & & & \\
\hline
\hline
\textit{Euclid}     & $(0.65,2.05)$ & $0.1$ & $15000$ & $0.001\ (1+z)$ & $0.76 \frac{\delta_{m}(\infty,z)}{\delta_{m}(\infty,0)}$ & $4.4 \cdot 10^{-3}$ & $3.5 \cdot 10^{-3}$ & $1.5 \cdot 10^{-3}$\\
\hline
\textit{WFirst-2.4} & $(1.,2.8)$    & $0.1$ & $2000$  & $0.001\ (1+z)$ & $0.76 \frac{\delta_{m}(\infty,z)}{\delta_{m}(\infty,0)}$ & $7.6 \cdot 10^{-3}$ & $6.7 \cdot 10^{-3}$ & $2.6 \cdot 10^{-3}$\\
\hline
\hline
\end{tabular}}
\caption{Surveys specifications. Column $1$: name of the survey; column $2$: redshift range; column $3$: redshift bin width; column $4$: survey area in sq. deg.; column $5$: redshift error; column $6$: galaxy bias functions; $k^{bin}_{max}$: largest wavenumber (smallest wavelength) from single bin volumes; $k^{bin}_{min}$: smallest wavenumber (largest wavelength) from single bin volumes; $k^{survey}_{min}$: smallest wavenumber (largest wavelength) from total survey volume.}\label{tab:surveys}
\end{minipage}
\end{table}}}
\end{center}

\subsection{Fisher Matrix calculation}

For the last step in the calculation of the Fisher Matrices elements, we need to define the non linear matter power spectrum. In a standard scale-independent context, it would have been defined, as usual, as:
\begin{equation}
\mathcal{P}_{NL}(k,z,\mu) = \left( b_{s}(k,z) + f_{s}(k,z)\mu^{2} \right)^{2} \frac{\mathcal{P}_{L}(k,z=0)}{\sigma^{2}_{8,0}}\, ,
\end{equation}
where: $\mathcal{P}_{L}$ is the linear matter power spectrum calculated by CAMB; $b_{s}(k,z) = b_{g}(z) \sigma_{8}(k,z)$; and $\mu = \frac{\overrightarrow{k} \cdot \overrightarrow{r}}{r}$ is the cosine of the angle of the wavenumber $\overrightarrow{k}$ with respect to the line-of-sight direction. But in this work, where we aim to highlight possible scale-dependence of the matter power spectrum, we will use a much more general expression, which should take into account for possible scale-dependence effects when approaching higher scales (i.e. smaller wavelength $k$). Following \citep{Jeong:2011as}, the non linear matter power spectrum will be defined as:
\begin{equation}
\mathcal{P}_{NL}(k,z,\mu) = \left[ \left( b_{s}(k,z) + f_{s}(k,z)\mu^{2} \right)^{2}  +
2 \left( b_{s}(k,z) + f_{s}(k,z)\mu^{2} \right) \frac{\mathcal{A}_{s}}{x^2} + \frac{\mathcal{A}_{s}^{2}}{x^{4}} + \mu^{2} \frac{\mathcal{B}_{s}^{2}}{x^{2}} \right] \frac{\mathcal{P}_{L}(k,z=0)}{\sigma^{2}_{8,0}}\, ,
\end{equation}
where
\begin{eqnarray}
x &=& \frac{k}{a H(a)}; \\
\mathcal{A}_{s}(k,a) &=& \sigma_{8,0} \mathcal{A}(k,a); \nonumber \\
\mathcal{A}(k,a) &=& \frac{3}{2} \Omega_{m} \left[ b_{e} \left( 1-\frac{2}{3} \frac{f(k,a)}{\Omega_{m,0}}\right) + 1 + 2\frac{f(k,a)}{\Omega_{m,0}} + \mathcal{C}(a) - f(k,a) - 2 \mathcal{Q}(a) \right]; \nonumber \\
\mathcal{B}_{s}(k,a) &=& \sigma_{8,0} \mathcal{B}(k,a); \nonumber \\
\mathcal{B}(k,a) &=& f(k,a) \left[ b_{e}(a) + \mathcal{C}(a) - 1 \right]; \nonumber \\
\mathcal{C}(a) &=& \frac{3}{2}\Omega_{m,0}(a) - \frac{1}{a H(a)} \frac{2}{D_{A}(a)/a} [1-\mathcal{Q}(a)] - 2 \mathcal{Q}(a); \nonumber \\
b_{e}(a) &=& \frac{d \ln \left( a^{3} \overline{n}_{g}(a) \right)}{d \ln a}; \nonumber \\
\mathcal{Q}(a) &=& -\frac{d \ln \overline{n}(>L,a)}{d \ln L}; \nonumber
\end{eqnarray}
with $\overline{n}_{g}$ the average galaxy density, which is given by each fiducial; and $\overline{n}(>L)$ is the cumulative luminosity function. This latter quantity could be recovered, for each survey we have considered, by checking details of the simulations which have been employed by each team, to obtain the nominal data discussed in the official papers. As we have checked that its value has basically no influence on our results, we have simply decided to fix the function $\mathcal{Q}$ to zero.

Finally, we calculate the observed matter power spectrum as:
\begin{equation}
\mathcal{P}_{obs}(k,z,\mu) = \mathcal{P}_{NL}(k,z,\mu) \exp \left[ -k^{2} (1-\mu^{2})\Sigma^{2}_{\perp}(z) -k^{2} \mu^{2}\Sigma^{2}_{||}(z) -k^{2} \mu^{2}\Sigma^{2}_{z}(z) \right] + \mathcal{P}_{shot}(z) \,,
\end{equation}
where the damping factors are needed to take into account the smearing due to non-linear structure formation \cite{Seo:2007ns} along $(\Sigma_{||})$ and across $(\Sigma_{\perp})$ the line of sight, and due to redshift errors $(\Sigma_z)$. Such damping factors are defined following \cite{Font-Ribera:2013rwa} as:
\begin{equation}
\Sigma_{\perp}(z) = dp(z) \cdot 9.4 \left[ \frac{\sigma_{8}(\infty,z)}{0.9}\right] \frac{1}{h}\,, \quad
\Sigma_{||}(z) = dp(z) \cdot [1+f(\infty,z)] \cdot \Sigma_{\perp}(z)\,, \quad
\Sigma_{z}(z) = \frac{c\, \sigma_{z}(z)}{H(z)}\,,
\end{equation}
where $dp(z)$ is an interpolating function defined in \cite{Font-Ribera:2013rwa} for the so-called ``$50\%$-reconstruction'', and $\sigma_{z}$ are the redshift errors. Note again that such functions are obtained in a context where the sub-Hubble and the slowly varying potential approximations are applied; thus, in order to use them properly, we consider the limit $k\rightarrow\infty$  for our solutions.

The elements of the Fisher Matrices are related to the derivatives of the observed matter power spectrum with respect to the variables of interest. We rewrite the non linear matter power spectrum considering the mapping from the fiducial cosmology to the real unknown cosmological background:
\begin{eqnarray}
\mathcal{P}_{NL}(k^{fid}_{\perp},k^{fid}_{||},z) &=& \alpha^{2}_{\perp}(z) \alpha_{||} \left\{ \left[ b_{s}(k,z) + f_{s}(k,z)
\left( \frac{k^{2,fid}_{||} \alpha^{2}_{||}}{k^{2,fid}_{||} \alpha^{2}_{||} + k^{2,fid}_{\perp} \alpha^{2}_{\perp}}\right)^{2} \right]^{2}  + \right. \\
&& \left. 2 \left[ b_{s}(k,z) + f_{s}(k,z)\left( \frac{k^{2,fid}_{||} \alpha^{2}_{||}}{k^{2,fid}_{||} \alpha^{2}_{||} + k^{2,fid}_{\perp} \alpha^{2}_{\perp}}\right)^{2} \right] \frac{\mathcal{A}_{s}}{x^2} + \frac{\mathcal{A}_{s}^{2}}{x^{4}} + \right.  \nonumber \\
&& \left. \left( \frac{k^{2,fid}_{||} \alpha^{2}_{||}}{k^{2,fid}_{||} \alpha^{2}_{||} + k^{2,fid}_{\perp} \alpha^{2}_{\perp}}\right)^{2} \frac{\mathcal{B}_{s}^{2}}{x^{2}} \right\} \frac{\mathcal{P}_{L}(k=\sqrt{k^{2,fid}_{||} \alpha^{2}_{||} + k^{2,fid}_{\perp} \alpha^{2}_{\perp}},z=0)}{\sigma^{2}_{8,0}}\, , \nonumber
\end{eqnarray}
where, following literature, we have defined:
\begin{equation}
\alpha_{\perp}(z) \equiv \frac{r^{fid}_{\perp}(z)}{r_{\perp}(z)} = \frac{D^{fid}_{A}(z)}{D_{A}(z)}\,, \qquad
\alpha_{||}(z) \equiv \frac{r^{fid}_{||}(z)}{r_{||}(z)} = \frac{H(z)}{H^{fid}(z)}\, .
\end{equation}
We have also used the simple geometrical definitions:
\begin{equation}
k^{2} = k^{2}_{||} + k^{2}_{\perp}\,, \qquad \mu^{2} = \frac{k^{2}_{||}}{k^{2}}\, ,
\end{equation}
and the transformation rules:
\begin{equation}
k_{\perp} = k^{fid}_{\perp} \alpha_{\perp}\,, \qquad k_{||} = k^{fid}_{||} \alpha_{||}\, .
\end{equation}
Given all these ingredients, a generic Fisher Matrix element, $F_{ij}$, with $i,j=1,\ldots,5$ (where $1$ stands for $\alpha_{\perp}$, $2$ for $\alpha_{||}$, $3$ for $f_{s}$, $4$ for $b_{s}$ and $5$ for the shot noise $\mathcal{P}_{shot}$), can be calculated as:
\begin{eqnarray}
      F_{ij}(z) &=& \frac{1}{4 \pi^{2}} \int^{1}_{-1} d\mu' \int^{k_{max}}_{k_{min}} dk' k'^{2}\frac{V_{eff}(k',z,\mu')}{2} \frac{d \ln \mathcal{P}_{obs}(k',z,\mu')}{d p_{i}} \frac{d \ln \mathcal{P}_{obs}(k',z,\mu')}{d p_{j}} \\
&\times&\exp \left[ -k'^{2} (1-\mu'^{2})\Sigma^{2}_{\perp}(z) -k'^{2} \mu'^{2}\Sigma^{2}_{||}(z) -k'^{2} \mu'^{2}\Sigma^{2}_{z}(z) \right]\,, \nonumber
\end{eqnarray}
where
\begin{equation}
k_{min} = \frac{2\pi}{V^{1/3}_{survey}}\,, \qquad k_{max} = 0.2\, h \,,
\end{equation}
and $V_{survey}(z)$ is the volume spanned by each survey in each redshift bin. Then, we calculate all the involved quantities, as the effective volume:
\begin{equation}
V_{eff}(k,z,\mu) = V_{survey}(z) \left( \frac{n(z) \mathcal{P}_{NL}(k,z,\mu)}{n(z) \mathcal{P}_{NL}(k,z,\mu) + 1} \right)^{2}\,,
\end{equation}
and the derivatives of the observed power spectrum with respect to al the variable we are interested. For sake of clarity and completeness, the interested reader can find all the derivatives required for the calculation in the appendix section.

\section{Results and conclusions}\label{lastS}

Results from the Fisher Matrix approach are reported in Table~\ref{tab:Fisher}, in which we present the errors expected from each reference survey case for the quantity related to the growth of the perturbations, which is actually measured from a galaxy survey, i.e. $f\sigma_{8}$. We present errors where no marginalization procedure on the Fisher Matrices is applied at all, i.e. we are assuming a realistic control/knowledge of all the details involved in a galaxy survey.

\begin{figure}
\begin{tabular}{cc}
  \resizebox{73mm}{!}{\includegraphics{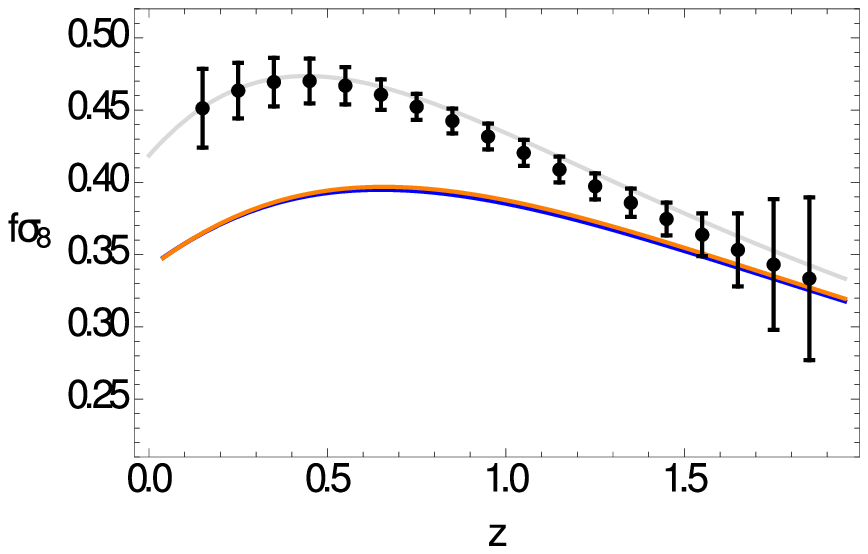}}&
  \resizebox{73mm}{!}{\includegraphics{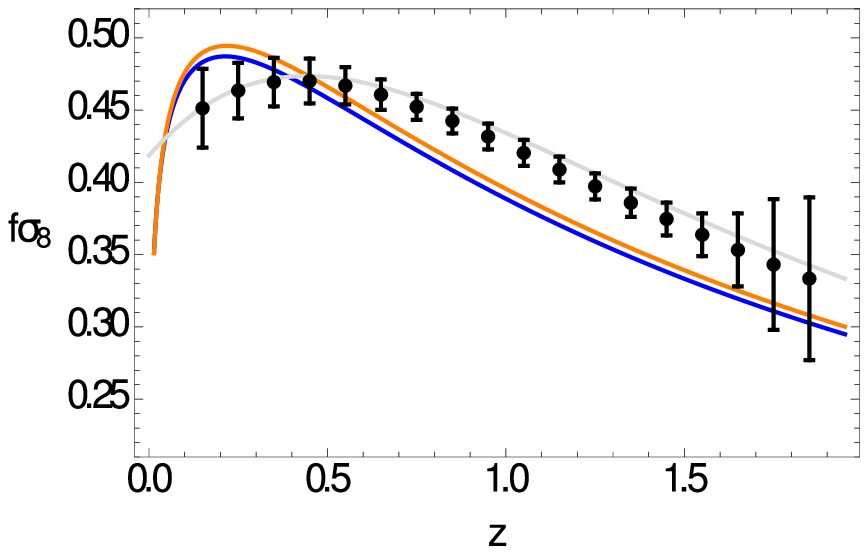}}\\
  \resizebox{73mm}{!}{\includegraphics{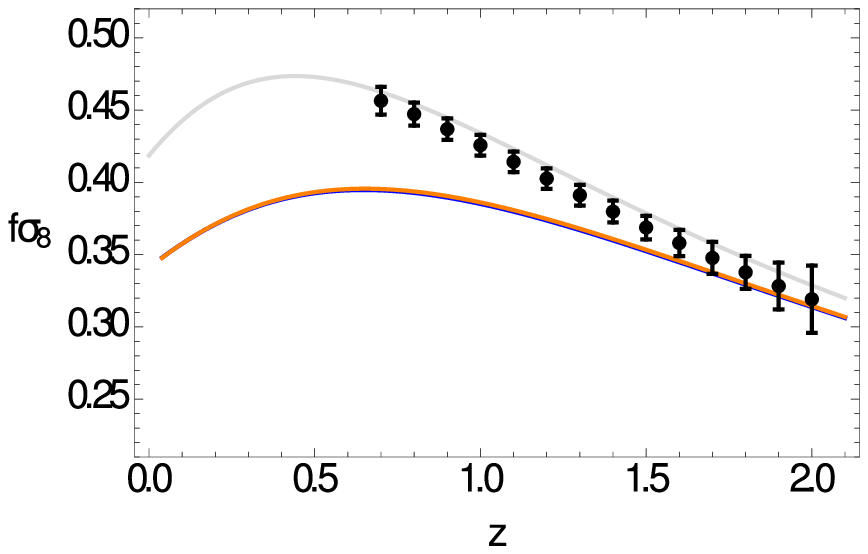}}&
  \resizebox{73mm}{!}{\includegraphics{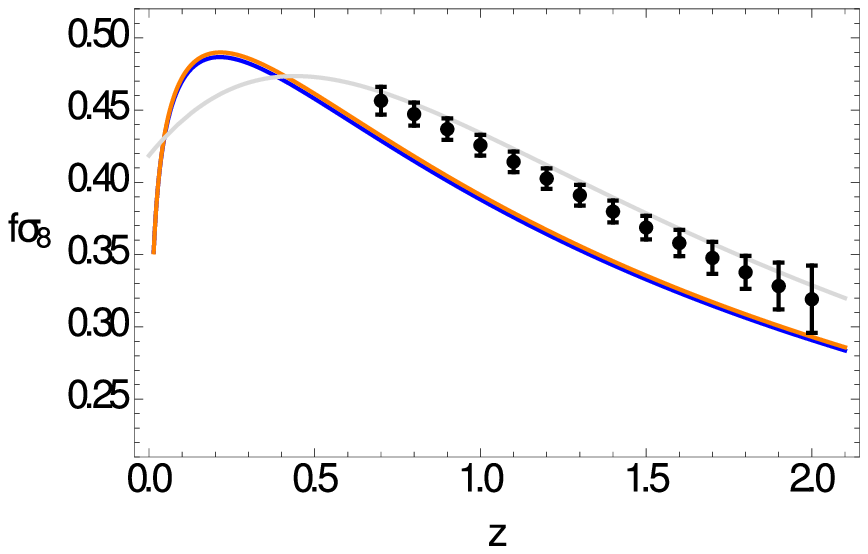}}\\
  \resizebox{73mm}{!}{\includegraphics{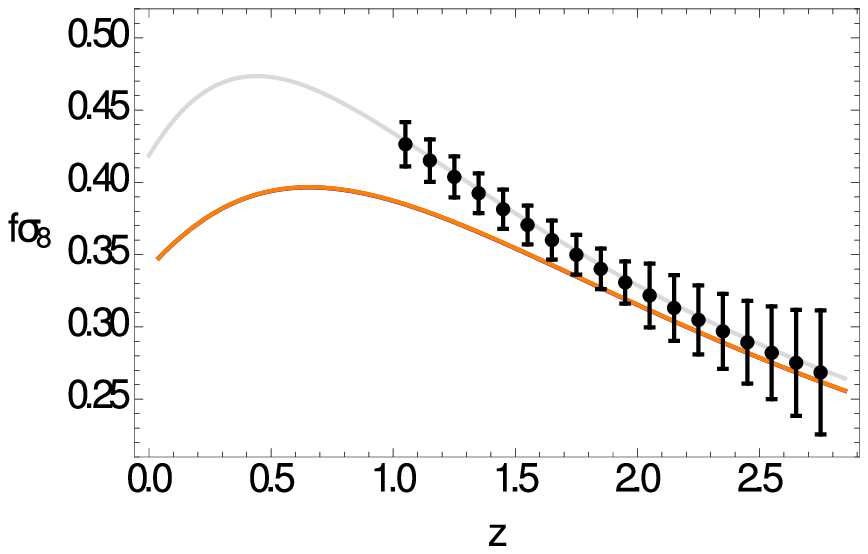}}&
  \resizebox{73mm}{!}{\includegraphics{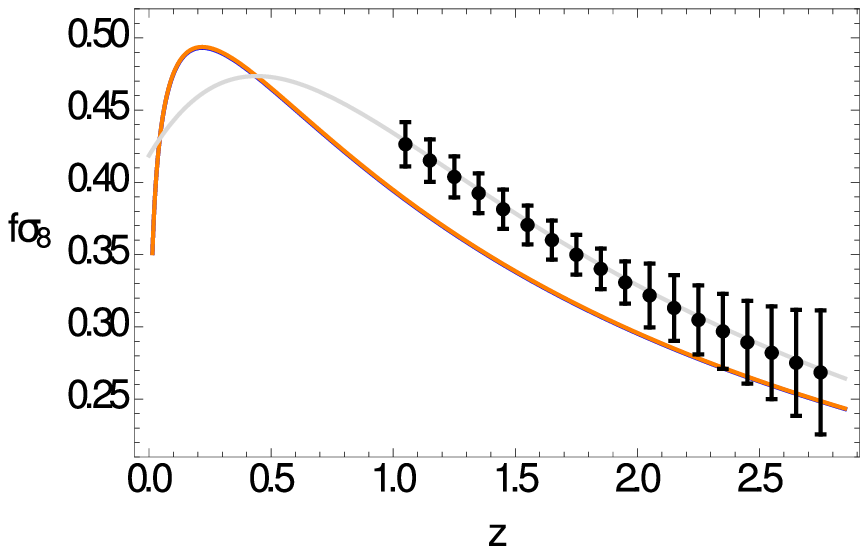}}
\end{tabular}
\caption{The plots of the $f\sigma_8$  for $\Lambda$CDM (grey line: scale-independent approach; black and grey points: scale-dependent approach), the SFS (left panel) and the FSFS (right panel) scenarios for DESI, \textit{Euclid} and \textit{W-First2.4}, from the top to the bottom respectively. Orange line is the scale-dependent solution assuming the smallest bin wavelength ($k^{bin}_{max}$); blue line corresponds to the scale-dependent solution assuming the largest bin wavelength ($k^{bin}_{min}$), as given in Table~\ref{tab:surveys}, for each survey. The error bars for the $\Lambda$CDM are given in Table~\ref{tab:Fisher}. The values of the parameters for the models with the future singularity are taken as for the SFS model: $m=0.749$, $n=1.99$, $b=-0.45$, $y_0=0.77$, for the FSFS: $m=2/3$, $n=0.7$, $b=0.24$, $y_0=0.96$. }\label{figura}
\end{figure}

In Fig.~\ref{figura} we plot the $f\sigma_8$ quantity, as a function of the redshift, which is obtained solving the full set of Eqs.~(\ref{gr1})~-~(\ref{gr3}). Solid grey line presents the $\Lambda$CDM fiducial model in the standard scale-independent approach; dots, instead, are also for the $\Lambda$CDM, but scale-dependent. Note, that each of the presented points has its own scale $k$, derived from the size of each redshift bin. In Table~\ref{tab:surveys} we only report the minimum $(k^{bin}_{max})$ and the maximum $(k^{bin}_{min})$ bin wavelength corresponding to each survey. Black points with errorbars represent the realistic error estimation from our Fisher Matrix procedure. For the SFS (left panel) and the FSFS (right panel) scenarios, we have fixed the model parameters at the values given in the caption of Fig.~\ref{figura}. Such values fulfil constraints given by the geometrical probes \textit{only}, i.e. SNeIa, BAO, CMB shift parameters and the Hubble data from passively evolving galaxies, as described in \cite{DHD,FSF,DDGH,rd,GHDD}. It is noteworthy to realize that the profiles which are shown in the figures, are from parameters which fit only geometrical probes and not the dynamical ones. Thus, they will be surely changed when such data will be taken into account (this issue will be dealt in a forthcoming work). For both models we only show the possible variability range of data as the continuous lines: the orange line corresponds to the scale-dependent case with $k^{bin}_{max}$; the blue line corresponds to the scale-dependent case with $k^{bin}_{min}$.

The first point to be addressed, is the way in which our forecast behave in the case of $\Lambda$CDM: both \textit{Euclid} and DESI will be able to discriminate between scale-independence and dependence in the redshift range $[0.75,1.65]$, even in the pessimistic scenario for error estimations. Instead, \textit{W-First2.4} won't be useful in this case. A visual inspection of the DESI and \textit{Euclid} surveys can also help us to discern what is the $k$-threshold (from bin volumes) where, approximately, the scale-independent and the scale-dependent approaches start to be different: we have that for $k\gtrsim0.005\, h$ Mpc$^{-1}$, we cannot detect any significant deviation, while for lower $k$ (i.e. larger scanned volumes/scales) we start to appreciate a more clear difference. In view of the above fact, it is clear why, for example, in the plot for \textit{W-First2.4} we can barely notice any discrepancy between the line for the scale-invariant and the scale-dependent $\Lambda$CDM data points: due to the small survey area, the bin wavenumber $k$ is always $>0.005\, h$ Mpc$^{-1}$ (see Table~\ref{tab:surveys}). On the contrary, for DESI, only low redshift bin will be more ambiguous, while higher redshift data will be much more straightforward. Finally, \textit{Euclid} will be optimal in this regard.

Comparing data from Table~\ref{tab:Fisher} with Fig.~(15) of \cite{Alam:2016hwk} we can also anticipate how significant will be the improvement in the accuracy of the data for future surveys: present data, as in Fig.~(15) of \cite{Alam:2016hwk}, span over a narrower redshift range and exhibit a larger dispersion, and both of these issues might be alleviated by performing observations within one single consistent survey. Moreover, the errors are $\sim\ 5$ times larger than what is expected from the future surveys.

For what concerns the power to distinguish among a $\Lambda$CDM and a singularity model, things are more murky, at least for one of the models we have chosen to work with. In fact, it is clear that the SFS model (left panel), as derived from geometric probes, is clearly not consistent with present data for the growth rate. In Fig.~(15) of \cite{Alam:2016hwk}, even taking into account the large dispersion of the data, and their small displacement toward smaller values of $f\sigma_8$ with respect to the \textit{Planck} $\Lambda$CDM cosmology, the data points are not distributed at values much lower than $f\sigma_8 \approx 0.4$. In the left panel of our Fig.\ref{figura}, instead, the SFS model never exceeds this value. Thus, we can expect, for the SFS model, a possible tension between the geometrical and the dynamical cosmological probes, but this will be topic for a forthcoming paper. About the FSFS model (right panel), we can see how it is, in principle, compatible with the data, and thus we expect not so large changes in its parameters even when taking into account dynamical probes. Moreover, it will be easy to rebut (or not) this model from future surveys: all the considered surveys can discard it at, at least, $3\sigma$, in the intermediate redshift range $[0.7,1.7]$.

All these considerations ``might be'' far more interesting in the scale-dependent approach to the evolution of the perturbations. The conditional is required in this case, because we can only infer a possible qualitative behaviour from the analysis we have at our disposal now. Nevertheless, the following considerations could serve as a good guide to set specifications of not yet planned surveys. Within this respect the case of DESI is very helpful. On one hand, the related, upper orange line corresponds to  $k=0.01\,h$ Mpc$^{-1}$, which is clearly higher than the threshold $k<0.005\,h$ Mpc$^{-1}$ that we have marked above and, thus, this line is practically equivalent to the scale invariant case. On the other hand, the relevant, lower blue line corresponds to $k=0.0036\, h$ Mpc$^{-1}$, which lies below our threshold. Thus, it corresponds to a regime in which, in the view of the size of the errors we have estimated, scale-dependence can be set precisely. The conclusion we draw from the above is that, while in the scale-invariant case, we would have observational data dispersed only around the orange line, thus making the discrimination between our models and the $\Lambda$CDM hard in the low-redshift regime, and more effective only in the high-redshift one, in the scale-dependent case we would have data dispersed in the region between the orange and the blue line which, thus, would exhibit a much more clear-cut deviation from the scale-independent $\Lambda$CDM case. And things should go even better with surveys spanning larger volumes or, within the same surveys we have considered so far, if larger redshift bins are considered, because in these cases we would reach values for $k$ much lower than the found threshold. However, comparing left and right panels, note also how the range span by the scale-dependent approach is not the same for each of the models we have considered, as if they were more or less sensitive to the scale-dependence. Thus, a possible model-dependence for it should also be taken into account.

In summary, in view of the new forthcoming data from galaxy surveys (not only by using BAO and redhisft space distortions, but also weak lensing) and the observations of the polarization of the CMB like CoRE and PRISM, it will be interesting to explore the details of the evolution of the perturbations in the dark sector. First, because having data for the growth of the perturbations, for the bigger wavelengths, and for the whichever redshift, we might be finally able to evaluate if worth to abandon the scale-invariant approach; second, because they will help to put more stringent constrains onto the class of the dynamical dark energy models with the singularities, like the sudden future singularity and the finite scale factor singularity scenarios, and any other dynamical dark energy or modified gravity scenarios.

\acknowledgments
We are grateful to Leandros Perivolaropoulos for his kind advice concerning the work. This project was financed by the Polish National Science Center Grant DEC-2012/06/A/ST2/00395.

\appendix
\section{Appendix: power spectrum derivatives}

In order to calculate Fisher Matrices, we need to calculate the derivative of the power spectrum with respect to some quantities. Here, we will list those we have used.

Derivative with respect to the shot noise:
\begin{equation}
\frac{d \ln \mathcal{P}_{obs}(k,z,\mu)}{d \mathcal{P}_{shot}} =  \frac{1}{\mathcal{P}_{obs}}\,.
\end{equation}
Derivative with respect to the growth rate:
\begin{eqnarray}
\frac{d \ln \mathcal{P}_{obs}(k,z,\mu)}{d f_{s}} &=& \frac{1}{\mathcal{P}_{obs}} \frac{\mathcal{P}_{L}(k,z=0)}{\sigma^{2}_{8,0}} \times \\
&&\left[ 2\mu^2\left( b_{s} + f_{s}\mu^{2}\right) + 2\mu^{2} \frac{\mathcal{A}_{s}}{x^2} + \left( \frac{2 \left( b_{s} + f_{s}\mu^2 \right)}{x^2} + \frac{2 \mathcal{A}_{s}}{x^{4}} \right) \frac{d \mathcal{A}_{s}}{d f_{s}} + \frac{2\mu^2 \mathcal{B}_{s}}{x^{2}} \frac{d \mathcal{B}_{s}}{d f_{s}} \right]; \nonumber \\
\frac{d \mathcal{A}_{s}(k,a)}{d f_{s}} &=& \frac{3}{2} \Omega_{m,0} \left[ -\frac{2}{3} \frac{b_{e}(a)}{\Omega_{m,0}} + \frac{2}{\Omega_{m,0}} - 1 \right]; \nonumber \\
\frac{d \mathcal{B}_{s}(k,a)}{d f_{s}} &=& b_{e}(a) + \mathcal{C}(a) -1. \nonumber
\end{eqnarray}
Derivative with respect to the galaxy bias:
\begin{equation}
\frac{d \ln \mathcal{P}_{obs}(k,z,\mu)}{d b_{s}} = \frac{1}{\mathcal{P}_{obs}} \frac{\mathcal{P}_{L}(k,z=0)}{\sigma^{2}_{8,0}} \left[ 2 \left( b_{s}(z)+ f_{s}(z)\mu^{2}\right) + \frac{2 \mathcal{A}_{s}}{x^{2}} \right].
\end{equation}
Derivative with respect to the BAO observables:
\begin{eqnarray}\label{eq:power_der}
\frac{d \ln \mathcal{P}_{obs}}{d \alpha_{\perp}} &=& \left( \frac{2}{\alpha_{\perp}} + \frac{d \ln \mathcal{P}_{L}}{d k} \frac{d k}{d \alpha_{\perp}} \right) \frac{\mathcal{P}_{NL}}{\mathcal{P}_{obs}} + \alpha^{2}_{\perp} \alpha_{||} \left\{ \frac{d \mu^2}{d \alpha_{\perp}} \times \right. \\
&& \left[ \left. 2 f_{s} \left( b_{s}+f_{s} \mu^{2} + \frac{\mathcal{A}_{s}}{x^{2}}\right) + \frac{\mathcal{B}^{2}_{s}}{x^{2}}\right] +
\left[ \frac{2\left( b_{s}+f_{s} \mu^{2}\right)}{x^{2}} + \frac{2 \mathcal{A}_{s}}{x^{4}}\right] \frac{d \mathcal{A}_{s}}{d \alpha_{\perp}} + \right. \nonumber \\
&& \left. \frac{2 \mu^2 \mathcal{B}_{s}}{x^{2}} \frac{d \mathcal{B}_{s}}{d \alpha_{\perp}} + \left[ - \frac{4\left( b_{s}+f_{s} \mu^{2} \right)\mathcal{A}_{s}}{x^{3}} - \frac{4\mathcal{A}_{s}}{x^{5}}- \frac{2\mu^{2}\mathcal{B}^{2}_{s}}{x^{3}}\right] \frac{d x}{d \alpha_{\perp}}\right\} \frac{1}{\mathcal{P}_{obs}} \frac{\mathcal{P}_{L}(k,z=0)}{\sigma^{2}_{8,0}}; \nonumber \\
\frac{d \ln \mathcal{P}_{obs}}{d \alpha_{||}} &=& \left( \frac{1}{\alpha_{||}} + \frac{d \ln \mathcal{P}_{L}}{d k} \frac{d k}{d \alpha_{||}} \right) \frac{\mathcal{P}_{NL}}{\mathcal{P}_{obs}} + \alpha^{2}_{\perp} \alpha_{||} \left\{ \frac{d \mu^2}{d \alpha_{||}} \times \right. \\
&& \left[ \left. 2 f_{s} \left( b_{s}+f_{s} \mu^{2} + \frac{\mathcal{A}_{s}}{x^{2}}\right) + \frac{\mathcal{B}^{2}_{s}}{x^{2}}\right] +
\left[ \frac{2\left( b_{s}+f_{s} \mu^{2}\right)}{x^{2}} + \frac{2 \mathcal{A}_{s}}{x^{4}}\right] \frac{d \mathcal{A}_{s}}{d \alpha_{||}} + \right. \nonumber \\
&& \left. \frac{2 \mu^2 \mathcal{B}_{s}}{x^{2}} \frac{d \mathcal{B}_{s}}{d \alpha_{||}} + \left[ - \frac{4\left( b_{s}+f_{s} \mu^{2} \right)\mathcal{A}_{s}}{x^{3}} - \frac{4\mathcal{A}_{s}}{x^{5}}- \frac{2\mu^{2}\mathcal{B}^{2}_{s}}{x^{3}}\right] \frac{d x}{d \alpha_{||}}\right\} \frac{1}{\mathcal{P}_{obs}} \frac{\mathcal{P}_{L}(k,z=0)}{\sigma^{2}_{8,0}}; \nonumber
\end{eqnarray}
where:
\begin{eqnarray}
\frac{d \mathcal{A}_{s}}{d \alpha_{\perp}} &=& \frac{3}{2} \Omega_{m,0} \sigma_{8,0} \frac{d \mathcal{C}}{d \alpha_{\perp}}; \\
\frac{d \mathcal{A}_{s}}{d \alpha_{||}} &=& \frac{3}{2} \Omega_{m,0} \sigma_{8,0} \frac{d \mathcal{C}}{d \alpha_{||}}; \nonumber \\
\frac{d \mathcal{B}_{s}}{d \alpha_{\perp}} &=& f_{s} \frac{d \mathcal{C}}{d \alpha_{\perp}}; \\
\frac{d \mathcal{B}_{s}}{d \alpha_{||}} &=& f_{s} \frac{d \mathcal{C}}{d \alpha_{||}}; \nonumber \\
\frac{d \mathcal{C}}{d \alpha_{\perp}} &=& -\frac{2\left( 1- \mathcal{Q}\right)}{\alpha_{||}}\frac{1}{H D_{A}}; \\
\frac{d \mathcal{C}}{d \alpha_{||}} &=& -\frac{3 \Omega_{m,0} H^{2}_{0}a^{-3}}{H^{2}} \frac{1}{\alpha^{3}_{||}} + \frac{2 \alpha_{\perp} \left( 1 - \mathcal{Q} \right)}{H D_{A}} \frac{1}{\alpha^{2}_{||}};  \nonumber \\
\frac{d x}{d \alpha_{\perp}} &=& \frac{c}{a \alpha_{||} H} \frac{d k}{d \alpha_{\perp}}; \\
\frac{d x}{d \alpha_{||}} &=& \frac{c}{a \alpha_{||} H} \frac{d k}{d \alpha_{||}} - \frac{k c}{a H} \frac{1}{\alpha^{2}_{||}}; \nonumber
\end{eqnarray}
\begin{eqnarray}
\frac{d \mu^2}{d \alpha_{\perp}} &=& -\frac{2}{\alpha_{\perp}}\mu^{2}(1-\mu^{2}); \\
\frac{d \mu^2}{d \alpha_{||}} &=& \frac{2}{\alpha_{||}}\mu^{2}(1-\mu^{2}); \nonumber \\
\frac{d k}{d \alpha_{\perp}} &=& \frac{k}{\alpha_{\perp}}(1-\mu^{2}); \\
\frac{d k}{d \alpha_{||}} &=& \frac{k}{\alpha_{||}}\mu^{2}. \nonumber
\end{eqnarray}
The derivatives of the linear power spectrum $\mathcal{P}_{L}$ with respect to the wavenumber $k$ are calculated numerically.

\begin{center}
{\renewcommand{\tabcolsep}{1.5mm}
{\renewcommand{\arraystretch}{1.5}
\begin{table}[ht!]
\begin{minipage}{\textwidth}
\centering
\resizebox*{0.4\textwidth}{!}{
\begin{tabular}{cc|cc|cc}
\hline
\hline
\multicolumn{2}{c|}{DESI} & \multicolumn{2}{c|}{\textit{Euclid}} & \multicolumn{2}{c}{\textit{WFirst-2.4}} \\
\hline
$z$ & $\sigma_{f\sigma_{8}}$ & $z$ & $\sigma_{f\sigma_{8}}$ & $z$ & $\sigma_{f\sigma_{8}}$ \\
 & $(\%)$ & & $(\%)$ & & $(\%)$\\
\hline
$0.15$  & $2.73$ & $-$ & $-$ & $-$ & $-$ \\
$0.25$  & $1.92$ & $-$ & $-$ & $-$ & $-$ \\
$0.35$  & $1.68$ & $-$ & $-$ & $-$ & $-$ \\
$0.45$  & $1.55$ & $-$ & $-$ & $-$ & $-$ \\
$0.55$  & $1.29$ & $-$ & $-$ & $-$ & $-$ \\
$0.65$  & $1.06$ & $-$ & $-$ & $-$ & $-$ \\
$0.75$  & $0.90$ & $0.7$ & $0.96$ & $-$ & $-$ \\
$0.85$  & $0.85$ & $0.8$ & $0.79$ & $-$ & $-$ \\
$0.95$  & $0.89$ & $0.9$ & $0.75$ & $-$ & $-$ \\
$1.05$  & $0.90$ & $1.0$ & $0.72$ & $1.05$ & $1.53$ \\
$1.15$  & $0.89$ & $1.1$ & $0.71$ & $1.15$ & $1.47$ \\
$1.25$  & $0.90$ & $1.2$ & $0.71$ & $1.25$ & $1.42$ \\
$1.35$  & $0.98$ & $1.3$ & $0.72$ & $1.35$ & $1.38$ \\
$1.45$  & $1.14$ & $1.4$ & $0.75$ & $1.45$ & $1.35$ \\
$1.55$  & $1.48$ & $1.5$ & $0.81$ & $1.55$ & $1.34$ \\
$1.65$  & $2.53$ & $1.6$ & $0.91$ & $1.65$ & $1.35$ \\
$1.75$  & $4.52$ & $1.7$ & $1.11$ & $1.75$ & $1.37$ \\
$1.85$  & $5.63$ & $1.8$ & $1.14$ & $1.85$ & $1.40$ \\
$-$     & $-$    & $1.9$ & $1.62$ & $1.95$ & $1.46$ \\
$-$     & $-$    & $2.0$ & $2.32$ & $2.05$ & $2.21$ \\
$-$     & $-$    & $-$    & $-$   & $2.15$ & $2.28$ \\
$-$     & $-$    & $-$    & $-$   & $2.25$ & $2.39$ \\
$-$     & $-$    & $-$    & $-$   & $2.35$ & $2.60$ \\
$-$     & $-$    & $-$    & $-$   & $2.45$ & $2.86$ \\
$-$     & $-$    & $-$    & $-$   & $2.55$ & $3.21$ \\
$-$     & $-$    & $-$    & $-$   & $2.65$ & $3.67$ \\
$-$     & $-$    & $-$    & $-$   & $2.75$ & $4.29$ \\
\hline
\hline
\end{tabular}}
\caption{Fisher Matrix results. Percentage errors on $f\sigma_{8}$ for the different surveys we have considered and described in the text.}\label{tab:Fisher}
\end{minipage}
\end{table}}}
\end{center}

\bibliography{growthscale3}{}

\end{document}